\documentstyle[12pt]{article}
 \input BoxedEPS
\SetRokickiEPSFSpecial  
\begin{document}
 \newcommand{\npb}[1]{Nucl.\ Phys.\ {\bf B#1}\ }
\newcommand{\prd}[1]{Phys.\ Rev.\ {\bf D#1}\ }
\newcommand{\prll}[3]{{\em Phys.\ Rev.\ Lett.} {\bf #1}(19#2) #3}
\newcommand{\nc}{\newcommand}
 \nc{\pl}[3]{ {\em Phys.\ Lett. }{\bf #1} (19#2) #3}
\nc{\PL}{\pl}
\nc{\np}[3]{ {\em Nucl.\ Phys. }{\bf #1} (19#2) #3}
\nc{\pr}[3]{ {\em Phys.\ Rev. }{\bf #1} (19#2) #3}
\nc{\beq}{\begin{equation}} \nc{\eeq}{\end{equation}}
\nc{\beqa}{\begin{eqnarray}} \nc{\eeqa}{\end{eqnarray}}
\nc{\lsim}{\begin{array}{c}\,\sim\vspace{-21pt}\\< \end{array}}
\nc{\gsim}{\begin{array}{c}\sim\vspace{-21pt}\\> \end{array}}

\begin{titlepage}

\begin{center}

\vspace{2cm}

{\hbox to\hsize{hep-ph/9612426 \hfill  MIT-CTP- 2591}}

 \vspace{2cm}

\bigskip

{\Large \bf  New Mechanisms of Gauge-Mediated Supersymmetry Breaking   }

\bigskip

  {\bf Lisa Randall}\footnote{Supported in part by DOE under cooperative
                 agreement \#DE-FC02-94ER40818,
NSF Young Investigator 
Award, 
Alfred P. Sloan
Foundation Fellowship, DOE Outstanding Junior Investigator 
Award.}\\

 \bigskip

 { \small \it   Center for Theoretical Physics\\
    Laboratory of Nuclear Science and Department of Physics\\
    Massachusetts Institute of Technology\\
Cambridge, MA 02139, USA \\}

  {\tt  lisa@ctptop.mit.edu } 

 \bigskip

 \bigskip

\vspace{2cm}

{\bf Abstract}
\end{center} 

We introduce new  mechanisms for  the communication
of supersymmetry breaking via gauge  interactions.  These 
models do not require  complicated
dynamics to induce a nonvanishing $F$ term
for a singlet. The first class of models  communicates supersymmetry breaking to the visible
sector through a  ``mediator" field  that transforms under both a messenger gauge
group of the dynamical supersymmetry breaking sector and the standard
model gauge group. This model has distinctive phenomenology;
in particular, the scalar superpartners should be heavier by at least
an order of magnitude than the gaugino superpartners. The
second class of models has phenomenology more similar to the ``standard" messenger
sectors. A singlet is incorporated, but the model does not require
complicated mechanisms to generate a singlet $F$ term. The role
of the singlet is  to couple fields from the dynamical symmetry
breaking sector to fields transforming under the standard model gauge group.  We also mention a potential solution to the $\mu$ problem.

  \end{titlepage}

\renewcommand{\thepage}{\arabic{page}}

\setcounter{page}{1}

\setcounter{equation}{0}

\baselineskip=18pt

\section{Introduction}

There are essentially two  ways to communicate 
  supersymmetry breaking. Supergravity mediated supersymmetry
breaking has the virtue of simplicity. Supersymmetry breaking
can occur in a so-called hidden sector, and no contortions
are required to transfer the breaking of supersymmetry
to the visible sector because it is automatically  accomplished via
Planck-suppressed
operators.\footnote{Even gravity-mediated
supersymmetry breaking  is complicated by
the necessity for a singlet $F$-term to give
a tree-level gaugino mass \cite{bkn}.}

Gauge-mediated supersymmetry breaking \cite{ads1,gm,dn,dns,dnns,pt} on the other
hand, conceivably
possesses several advantages. Probably the most important from a
phenomenological
 perspective is that flavor changing neutral currents
are naturally suppressed.  It is perhaps also somewhat  more comfortable
to have the physics of supersymmetry at lower energy scales,
although the choice is nature's and not ours.

So far however, gauge-mediated supersymmetry breaking appears
not  to be  as attractive an  option as one would like.  Although
models for communicating supersymmetry breaking exist \cite{dn,dns,dnns}, they are quite
cumbersome. Furthermore, they generally have potentially
dangerous color breaking vacua \cite{we}. From our perspective, the
first problem is the more serious complaint, since it is hard
to believe that nature has chosen the complicated mechanisms
which are currently discussed in the literature.

There are several reasons why communicating supersymmetry
breaking via perturbative interactions at a low-energy scale appears
to require complicated structure.  It is commonly accepted that the ideal
scenario
would somehow embed the standard model gauge group into
a dynamical supersymmetry breaking sector in such a way
that standard model particle superpartners automatically have
the requisite supersymmetry breaking mass. However, the
generation of such a model
has   proved difficult. The primary problem is that to obtain
all the required soft supersymmetry breaking masses (including the gluino),
one generally runs
into a problem with a low-energy Landau pole due to the
large number of flavors carrying standard model gauge charge \cite{ads1}.  

For this reason,
the idea of a messenger sector was introduced, which  less directly
communicates the breaking of supersymmetry to the visible sector \cite{dn}.  
A messenger sector includes a vector representation
of the standard model. The most popular of these models
couples this vector representation to a singlet \cite{dns,dnns}
which has a nonzero $A$ and $F$ component, thereby
transmitting supersymmetry breaking to the messengers,
and consequently, to the visible sector.  For example,
a gluino mass is generated by the diagram of Figure 1,
which clearly requires  both  an $A$ and $F$ type VEV 
  in order to flip chirality on both the fermion and scalar
lines.
  This
is readily seen explicitly, or by considering  the U(1) carried by the vector
quarks.

\begin{figure}
$$\BoxedEPSF{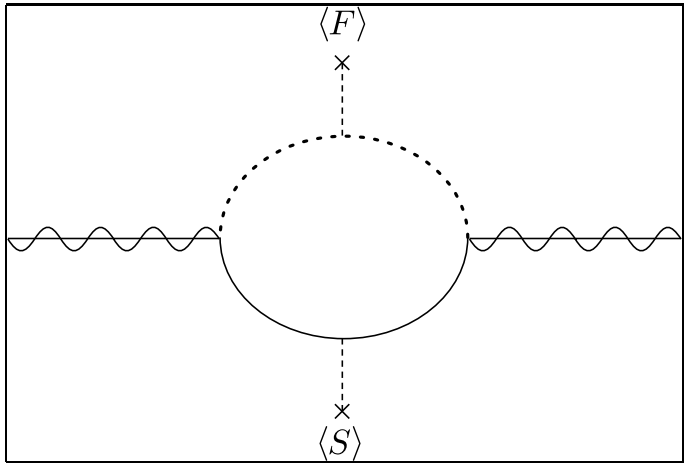 scaled 600}$$
 \caption{Conventional diagram to generate a gaugino mass.}
\label{fig:F1}
\end{figure}

There are two beneficial features of a messenger sector,
in addition to the eradication of the dangerous Landau pole.
First is that the gluino diagram is generated at one loop,
while the squark mass {\it squared} is generated at two loops,
so that the resulting squark and gluino mass are of the same
order of magnitude.\footnote{I will refer to squarks and gluinos
explicilty; corresponding results apply to sleptons and charginos, etc.}
Furthermore, the gaugino mass requires a single insertion of a supersymmetry-breaking $F$-term, while the
squark mass squared   requires two. If there is a singlet VEV, $S$,
which  generates the messenger fermion mass, 
the gaugino mass is of order $F/S$ and  the squark mass squared is
of order $(F/S)^2$, again a desirable relation. 
Together, these relations guarantee that the squark and gluino
masses are about the same. The second
advantage to models in the literature is that an explicit computation of the relevant
Feynman diagrams \cite{dn,martin,apt} shows that the
squark mass squared arising from the single $F$ term in the hidden
sector is positive.

Despite these advantages, one might  nonetheless reserve 
enthusiasm for   these models, primarily because of the complications
which are employed to give the singlet an $F$-type VEV. Singlets
generally do not play a role in dynamical symmetry breaking models.
A complicated scenario is  generally required to communicate
supersymmetry breaking to the singlet coupled to the messenger
sector.
 When a fundamental
singlet is coupled directly into
the dynamical symmetry breaking sector, it  appears  to be
 nontrivial to  couple
the singlet to messenger quarks without introducing
a flat direction in which    squarks get a VEV.
It is the generation of the $F$-type insertion for the singlet that seems
to be the key problem for generating gauge-mediated models
of supersymmetry breaking.  

A nice alternative to fundamental singlets was proposed by Poppitz
and Trivedi \cite{pt}, in which the messenger sector is embedded
directly into 
 the supersymmetry breaking sector. 
The usual problem with the Landau pole is avoided
because of the existence of two scales. Although
many states carry standard model quantum numbers,
many are heavy. The low energy sigma model
can be chosen to avoid a bad Landau pole.    

Although it is more compelling to have a model in which
singlets are not introduced ``by hand", these
models are in practice problematical. The first
problem is that it is in fact nontrivial, though not impossible
\cite{mrm} to embed
a group as large as the standard model as a global
symmetry group into the dynamical symmetry breaking sector.
The second problem is that in models with a hierarchy of
scales, there is necessarily a mass range for which $Str(M^2)\ne 0$
where the supertrace is taken over the messenger fields.
Explicit calculation \cite{martin,apt}  for existing
models in which $Str(M^2)>0$ shows
that this scenario generally implies {\it negative} mass
squared for the squarks, unless parameters and models
are carefully chosen. 

My goal in this paper is to explore alternatives for
communicating supersymmetry-breaking via gauge interactions
which do not require an $F$-term for a singlet which
couples directly to messenger quarks. 
 In the  models I present, there is a different paradigm
for simplicity than the first one suggested,
in which the gauge-mediated model  embeds
the standard model within the supersymmetry breaking sector.
The paradigm is more similar to that suggested by
 hidden sector models,   in which supersymmetry breaking can be communicated to
other sectors in a fairly generic fashion, without  structure which
relies on a particular hidden sector model.    In this paper  we  try to see how
far we can get with simpler structure to the messenger sector
and suggest  examples. The major distinguishing
feature of our models is that we permit there to be tree-level
mass terms in the superpotential. Naively, this might
seem to be counter to the philosophy of dynamical models,
in which one avoids introducing mass scales by hand. However,
we have learned in recent years that generating mass terms
dynamically is very straightforward. Simple examples
arise from compositeness, or dimension three Yukawa couplings
matching onto mass terms due to strong dynamics \cite{yuk}.  A mass term
could even be generated more prosaically from a Yukawa
coupling to a field which obtains a mass because    a mass squared
scales negative. The only real requirement we would like to  impose
is that it is not necessary to take two different mass scales with
entirely separate origin to be the same. This would amount to fine-tuning,
and is the reason we   believe the $\mu$-term requires some further
explanation. In our theories, there will be qualitative requirements
on mass parameters (that they be large or small compared to other
masses) but  different mass scales  are not required to coincide.

The first class of models, discussed in Section 2, requires
a messenger gauge group and  ``mediator" quarks which transform
  both under the messenger and standard model gauge groups.
 The squark and gluino masses both arise at high loop order,
in such a way that the squark is generically predicted to
be heavier than the gluino by at least an order of magnitude.
This would imply a relatively light gaugino (or heavy scalar) spectrum, subject to experimental
verification. 
The mediator models  can employ dynamical
models of a messenger sector,  in a way in which
both problems  mentioned above are solved.

In the second class of models, discussed
in Section 3,  I incorporate a heavy singlet ``intermediary"
field. The main distinguishing feature of this class of theories from
other gauge-mediated models with singlets 
is the fact that the singlet does not acquire an $F$ term via complicated
interactions
and no messenger gauge group is necessary. The singlet is present
 in order to  generate a higher dimension operator
which connects the symmetry breaking sector to the messenger sector.

We briefly discuss phenomenology in Section 4.
We conclude in the final section.   An Appendix gives examples
of dynamical supersymmetry breaking sectors which can
be used in mediator models.
 
\section{ Mediator Models}

The first class of models is based on the fact that  a gluino mass
can be generated by the diagram of Figure 2, instead of that
of Figure 1.\footnote{ A similar diagram was considered
in a somewhat different context in Ref. \cite{bogdan}.}   (Of course
one must also include the supersymmetric
 analogs; we present explicitly only
the diagram in which supersymmetry breaking
is communicated.) The necessary fields and interactions
for such a diagram to exist are the following.

 \begin{figure}
$$\BoxedEPSF{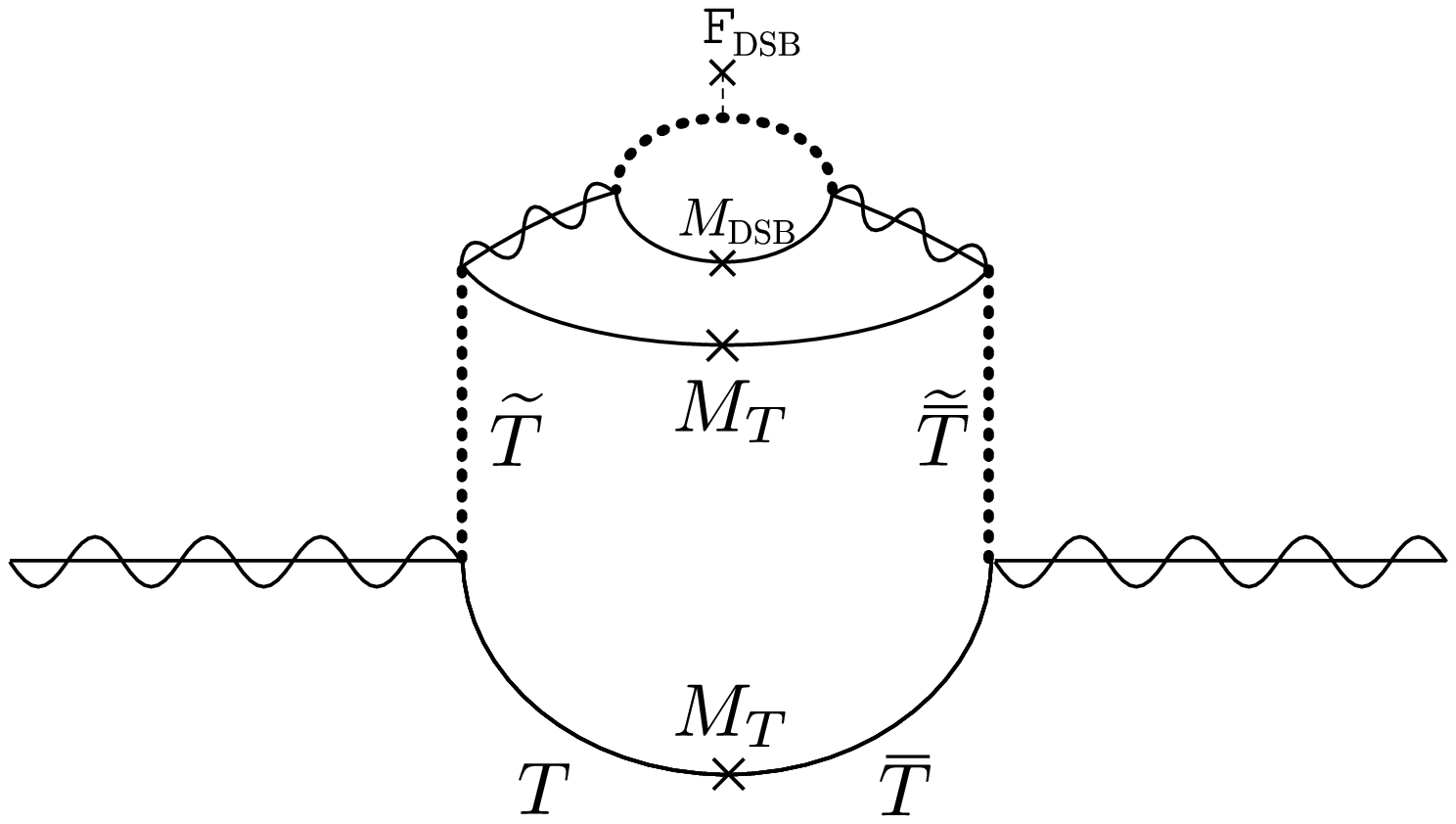 scaled 600}$$
 \caption{New mechanism for generating a visible sector gaugino mass. $T$ and $\overline T$
denote
fermions, whereas $\widetilde T$ and $\widetilde{\overline T}$ denote scalars.}
\label{fig:F2}
\end{figure}

There is a weakly gauged global symmetry  $G_m$ acting
on fields in the dynamical supersymmetry breaking sector.
This gauge group may or may not be broken, but a supersymmetry
breaking gaugino mass, $M_m$, for the gauge bosons of this group
is essential.  The existence of the messenger
gauge group, and the existence of a supersymmetry
breaking messenger gaugino mass, are the aspects
of the model which depend on the dynamical symmetry
breaking sector. Examples of models with these
properties are discussed in  an Appendix. Second, there 
are ``mediator fields", which we call $T$ and $\bar{T}$, in  a vector representation
of the standard model,   which transform
both under the messenger gauge group, $G_m$, and the standard
model gauge group, $G_{SM}$ (or an extension thereof)
(and therefore ``talk" to both the dynamical supersymmetry
breaking and visible sectors).  In order
to keep the number of flavors of the standard model small,
we will restrict our attention here to $G_m$  having
rank up to four, SU(2) being probably
the simplest group to accomodate.  For example,  $T$ can transform under $SU(2)_m\times
SU(5)_{GUT}$ as a $(2,5)$ and $\bar{T}$ as a $(2,\bar{5})$.
   Note that grand unification is not
essential and the $T$'s can transform under the
standard model gauge group, $SU(3)\times SU(2) \times U(1)$ instead. Third,
there is a supersymmetric
mass term $M_T T \bar{T}$. This might seem counter to the philosophy
of dynamical models, in which mass scales are not specified by hand.
The real requirement however is that the mass scales should
not be introduced in a fine-tuned fashion. We will see
shortly that $M_T$ is constrained to   lie between $ M_m$,
the messenger gaugino mass, 
and the mass of the heaviest of the  hidden sector scalars, $M_{DSB}$. If there is no large hierarchy then $M_T$ must
accidentally agree with the scales in the supersymmetry
breaking sector. If there is a large hierarchy between
the above scales, as exists in
some calculable models of supersymmetry breaking, or when
supersymmetry is broken through higher dimensional operators,
there can be a substantial range for $M_T$ and the model
is  natural. 

With these ingredients, supersymmetry breaking can
be communicated to the visible sector via the $T$ field
so long as there can be a supersymmetry breaking $G_m$
gaugino mass.  Examples where the messenger
gaugino will be massive are discussed
in  an Appendix.
In this model, the gluino and squark mass are both generated at high 
loop level. It should be noted that  the complicated Feynman diagrams
do not make the theory more complicated; they are present
whether or not we compute them.  However, because
these are higher loop diagrams,  
we estimate, rather than  compute the masses of the superpartners
of the visible sector.  This is sufficient
to allow us to identify the dependence on couplings,
loop factors, logarithms, and power dependence
on masses.   For this estimate,  it is most expedient to divide
the analysis into three possible ranges of parameters 
according to the  relative sizes of $M_T$  and $M_{DSB}$, where $M_{DSB}$
is the mass of the heaviest field with a supersymmetry breaking
scalar mass.

{\bf $M_T    \sim  M_{DSB}$:} This mass range is not necessarily
natural, as $M_T$ and $M_{DSB}$ have separate origin, counter
to the philosphy espoused in the introduction. We include
it for three reasons. First, pedagogically, it is simplest to
first count loop factors, independently of mass suppressions
or enhancements, which can be done most simply when all
masses are comparable.   Second, it is a logical possibility
which merits consideration. Third, even though we
do not know the solution to the $\mu$ problem, there
must be one; whatever  mechanism relates the scale $\mu$ to the
scale of soft supersymmetry breaking can  in principle  
 relate the $T$ mass to $M_{DSB}$.   

The gluino mass is generated by the Feynman diagram of Figure ~2.
  The gluino mass  is nonzero, even without the
presence of an $F$-term coupled directly to the messenger
squarks $T$ and $\bar{T}$. This is because the two-loop subdiagram
which has external $\tilde{T}$ and $\tilde{\bar{T}}$ states
plays the role of $F_S$ in completing the Feynman diagram.
The gaugino mass evaluates to a number of order $\alpha_s \alpha_m^2
F_{DSB}/(4 \pi)^3 M_{DSB}$ as it occurs at three loops.  Here
we have used the fact that $G_m$
is a weakly coupled gauge group which does not play
an essential role in the supersymmetry breaking dynamics.
Therefore, $M_m$ (really  the innermost loop
of Figure ~2 which is a self-energy diagram for the $G_m$ gauge boson) is  loop-suppressed, and is  of order   $ \alpha_m  F_{DSB}/4 \pi M_{DSB}$ 
(for this parameter regime there is not necessarily a distinction
between these two mass scales) and is generated in the usual way the gluino mass is generated in visible sector
models.

 The squark mass squared is generated by four-loop
diagrams not explicitly shown.   Some of these four-loop diagrams
can 
be readily  identified by inserting a two-loop diagram of
the standard sort with intermediate
DSB states \cite{dn,martin}
 that generates a supersymmetry breaking $\tilde{T}$ or $\tilde{\bar{T}}$
mass into a two-loop diagram with intermediate $T$ states
 that generates the squark mass squared.
 The squark
mass   evaluates to a number of order $\alpha_s \alpha_m
F_{DSB}/(4 \pi)^2 M_{DSB}$. The sign of the mass squared
is not known    without an explicit calculation, but
we expect one can choose parameters for which
it is positive. 

There also exist contributions to the $T$ squark mass which
arise by inserting the effective $``F"$ term which is generated
at two loops. However, these contributions give rise
to a six-loop contribution to the squark mass squared and are
therefore negligible. This can be seen also from the form
of the $T$ squark mass matrix.  In addition to the diagonal
supersymmetry breaking contribution to the $ \delta M_{\tilde{T}}^2 \tilde{T}^* \tilde{T}$
mass there is the off-diagonal $m^2_{LR} \tilde{T} \tilde{\bar{T}}$
type mass. These mass squared parameters are of the same
order of magnitude. However,    $m^2_{LR}$ appears
squared, whereas $\delta M_{\tilde{T}}^2$ appears once  in generating
a squark mass squared. Therefore the off-diagonal contribution
is negligible and can be neglected.

 We conclude that the ratio of  gluino to squark
mass is suppressed by $\alpha_m/(4 \pi)$ in this model.
 The question  then  is  what
is a reasonable range
for $\alpha_m$.
 The best situation is if $\alpha_m$ is  as large as possible
so that one does not run into naturalness problems. \footnote{I
thank Bogdan Dobrescu for stressing the viability of large $\alpha_m$.}
For $\alpha_m$ about 1,  in which case $G_m$ can still be considered
weakly coupled, this ratio is about 10, which is probably acceptable,
  and points to interesting predictions.
  If $G_m$ is not
asymptotically free, a dangerous Landau pole will develop  if $g_m$ is
much bigger than 1. On the other hand, $G_m$ can have rank up to
three or four, without running into problems with the running of SU(5).
Even for SU(2), with six flavors of doublets (the minimum content
required to have five flavors of messengers plus a doublet  flavor in the DSB sector), the  coupling does not run at leading order. So it is reasonable
to expect that the messenger group can allow  fairly big values of $\alpha_m$.\footnote
{$G_m$ should not be a $U(1)$ gauge group
in any case, because the presence of the $T$ fields
which transform under $U(1)$ and $U(1)_Y$ can
generate dangerous kinetic energy mixing terms \cite{jmr}.} The other potential problem with large $\alpha_m$  
is that the analysis of the supersymmetry breaking
vacuum might need to incorporate the extra gauge group,
and the separation  between the DSB and messenger
sectors would not be as clean. However the
model would still break supersymmetry, since the equations
of motion would still be inconsistent. Whether supersymmetry
is broken should be independent of the ratio of couplings
in the theory; if supersymmetry is broken at weak coupling
it should also be broken when the coupling is somewhat bigger.

  We conclude that  there is a hierarchy between the gaugino
and scalar masses (of like charges) of order 10.  This
prediction will persist for the other acceptable parameter
range $M_T<M_{DSB}$ as we will see shortly.
The  precise
ratio of scalar to gaugino mass depends on numbers expected to be of order unity which
we have not  incorporated. It is almost certainly a prediction of our
model that the gauginos will be   light and the scalars relatively heavy. 
We briefly discuss existing bounds in Section 4.   To determine the naturalness
of this parameter range also requires   assumptions about the  $\mu$
term and a more detailed calculation of the relevant Feynman diagrams.
   We
interpret the large mass ratio as a prediction of our class of theories
to be tested at future colliders. A more detailed analysis of the
viable parameter range would be very worthwhile.

{\bf $M_T>M_{DSB}$:} We next consider the possibility that
$M_T$ is the heaviest mass scale. We now
show that this possibility is not viable, because the gaugino
mass would be  suppressed by mass as well as loop factors
in comparison to the
squark mass, and is therefore too small.

This can be seen by an operator analysis, by explicit evaluation
of the Feynman diagrams, or by an effective theory
calculation of the Feynman diagrams.

Let us first consider the operator analysis.\footnote{I  am grateful to 
Ann Nelson for    discussions.}  In terms of a spurion
whose $F$-component breaks supersymmetry ($\Phi$) and 
which we also assume to have a nonvanishing $A$-component,
we can construct the operators:
 \beq \label{ve} 
\int d^4 \theta {  {Q}^\dagger {Q} \Phi^\dagger \Phi \over M_T^2}
\eeq
which gives rise to the squark mass squared and the operator
\beq \label{vev}
\int d^2 \theta {W_\alpha W^\alpha \Phi \Phi \over M_T^2}
\eeq
which contributes to the gluino mass.  The two factors
of $\Phi$ are required because we need both an $A$ and
$F$-type insertion of $\Phi$ in the relevant Feynman-diagram.
 From the above, it is readily concluded that the squark
mass  goes like  $F/M_{T}$ where as the gaugino mass  
is suppressed by  $F/(M_T)^2$. The additional suppression by $M_T$ is
undesirable since the gaugino mass is already light. In fact,
there are larger contributions to the squark mass squared,
not at all suppressed by $M_T$, so the ratio is even
worse than is suggested by the operator analysis. This
is understood from Feynman diagrams directly, or
by including operators involving a nonvanishing $Str$.
We briefly discuss the Feynman diagrams here, and
reserve the operator analysis with nonvanishing $Str$ to
the third case, where it is of more phenomenological relevance.

The  $M_T$  suppression  of the gluino mass can be readily understood from
the Feynman diagram directly.  The
innermost loop can be thought of as generating a gaugino
mass $M_m$  if the momenta running in the loop  are less
than $M_{DSB}$. However, because the relevant momenta
from the full diagram are of order $M_T$, the result
is suppressed by $F_{DSB} M_{DSB}/M_T^2$, where the
first two factors were necessary to complete the innermost
loop of Figure ~2.  This can also be seen from the effective
theory below the $T$ mass scale. There is a two-loop diagram
which generates  an operator suppressed by two powers of $M_T$.
Closing the loop of the light DSB fields again gives the suppression
factor we have discussed.    

One can also study the Feynman diagrams contributing
to the squark mass squared. In fact one finds that
the operator analysis above  misses large contributions
  to the squark mass which are not
mass suppressed at all. These can be understood
as    the $T$ scalar
 having an unsuppressed contribution to its mass splitting
which does not decouple when computing the squark mass squared.
(Further discussion of such effects are in the following section.)

We conclude that this case is not interesting from the point
of view of gauge-mediated models.

{\bf $M_m<M_T<M_{DSB}$} For this range to exist requires
that there is a hierarchy between the scales $\sqrt{F_{DSB}}$ 
and the scale $M_{DSB}$, the scale of  mass for the heaviest multiplet
in the DSB sector with a nonsupersymmetric realization. Many
calculable models of supersymmetry breaking have such a hierarchy,
as do models in which supersymmetry breaking occurs through
the presence of nonrenormalizable operators in the superpotential.
This is perhaps the most natural regime for our models.

The first point to understand is that the ratio of gluino to squark mass will not
be suppressed by positive powers of
 $M_T$. Naively, this might have seemed
to be the case because the gluino mass requires nonzero $M_T$
whereas the squark mass does not. 

We can analyze the gluino mass in this model most simply,
because we first integrate out the massive hidden sector
fields (those with mass greater than $M_T$). This
simply gives the standard one-loop messenger gauge boson
mass $M_m$ which can be considered ``hard" below the scale $M_{HS}$.
So the diagram contributing to the gluino mass is simply two-loop,
as in Figure ~3. The first inner loop generates a mass term
of the form $m^2_{LR} \tilde{T} \tilde{\bar{T}}$, where
$m^2_{LR} \propto M_m M_T$ is generated by the inner loop.
When inserted into the final loop, which is infrared convergent,
the factor of $M_T$  from $m^2_{LR}$ is cancelled by an
$M_T$ from the remaining loop   
(really one in the numerator divided by two in the denominator)
to give a result which is independent of $M_T$ and
depends on the same mass scale as the squark with no further
mass suppression factors.  This results holds only insofar as
$M_T>M_m$. For $M_T$ smaller than $M_m$, one obtains
the necessary mass suppression factor which yields a vanishing
result when $M_T\to 0$.

\begin{figure}
$$\BoxedEPSF{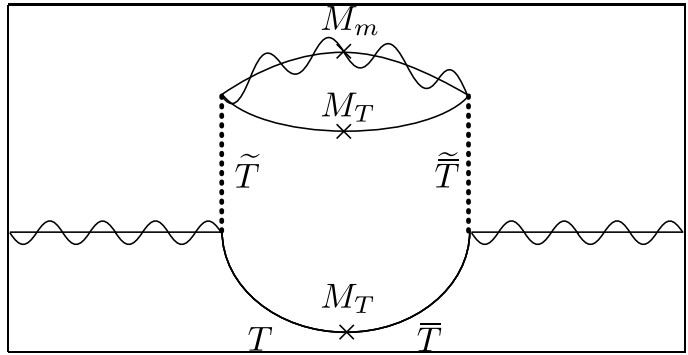 scaled 600}$$
 \caption{ Two-loop diagram which generates the gluino mass 
when $M_{DSB}\gg M_T$  }
\label{fig:F3}
\end{figure}

The two-loop diagram which gives the gluino mass (once $M_m$
is considered ``hard") has the interesting feature that
there is a logarithmic enhancement of the result due
to the divergent inner loop. The result for the gluino
mass is therefore of the order of $\alpha_m^2 \alpha_s Log(M_T/M_{DSB})^2 F_{DSB}/(4 \pi)^3 M_{DSB}$.

  The four-loop  calculation of the  squark mass is more
difficult and subtle.\footnote{I am grateful to Erich Poppitz
for discussions, and for sharing the results of \cite{apt} before
publication.}
Here it should be borne in mind that there are two scales of
mass for hidden sector fields. For example, in the model of Poppitz
and Trivedi \cite{pt}, briefly discussed in  an Appendix,
the mass of the light fields in the low-energy sigma model
is of order $\sqrt{F_{DSB}}$, whereas there are also
heavy fields with mass scale set by $M_{DSB}$ (times a gauge
or Yukawa coupling).  The  heavy fields couple
to light fields with  nonvanishing $F$-components, and  contribute to  both gaugino and squark masses.
However, the diagrams with internal light fields for DSB states
(lighter than $M_T$) also contribute to the squark mass, without
mass suppression! This seems to contradict the wisdom of
the operator analysis which suggested that  for $M_T>M_{DSB}$,
the Feynman diagrams contributing to the squark mass
squared would decouple. The essential difference
is that  the previous discussion only applies to  $F$-type
contributions to the mass squared.  In models
with a separation of scales, there is very likely  also a contribution
from nonvanishing  $Str(M^2)$      between
the two mass scales of the DSB scalars. This is in
fact the case in the models of Ref. \cite{pt} for example.
And the $T$ sector always has nonvanishing $Str (M^2)$.
   The mass for the $T$ scalar contributes in turn
to the squark mass squared, and  for this mass
range should be the  dominant
contribution  since it is logarithmically enhanced. Before
examining this contribution in more detail, let us first
understand the nondecoupling of the $Str(M^2)$ term.

Before we argued that when $M_T$ is large, the
squark mass squared will be suppressed by $F_{DSB}^\dagger F_{DSB}$.
Because this is dimension four, the dimensions are compensated
by two factors of $M_T$, yielding the mass suppressed results
we found earlier. However, $Str(M^2)$ is dimension two. If
$Str(M^2)$ has magnitude of order $F_{DSB}$ (as it does in Ref. \cite{pt}),
it
can be thought of as $F_{DSB}^\dagger F_{DSB}/|F_{DSB}| $,
where the dimensionful factors in the denominator are already
supplied, and are the lighter scale  $|F_{DSB}|$.   So whereas the contributions
due to $F$-type insertions decouple  when inserted
on a  heavy  internal line, this
is not true for the contributions proportional to $Str(M^2)$.
Futhermore, the  diagrams in which $F$-terms are not inserted
on the heavy line do not decouple, as can be seen by routing
the momentum on the two-loop diagram contributing to the $T$
mass so that there is no common loop momentum between the $T$
line and the line on which the $F$ terms are inserted. The
fact that $Str$ insertions do not decouple when inserted
on a heavy line whereas $F$ type insertions do
  can be readily seen from a one-loop computation (simpler than the  two- or four-loop calculations relevant to this model). Consider
  a quartic interaction between two heavy fields (mass $M_H$)
and two light fields, in which there
are both supersymmetry  breaking $F$-type
mass for heavy fields $\tilde{H}$ and $\widetilde{\bar{H}}$ and supersymmetry
breaking $\delta M^2$ masses for $\tilde{H}$ and $\tilde{\bar{H}}$.
 One can calculate
the mass squared for the light fields obtained by closing the heavy
fields in a loop and inserting $F$ twice. There are three
propagators and the result is suppressed by $M_{H}^2$. On
the other hand, when one calculates the contribution from
inserting $\delta M^2$, there are only two propagators--the
result is not only unsuppressed; it is divergent, and generates
a logarithmic contribution to the running
of the scalar mass squared for the external field.  These results are nothing new;
they are the reason gauge-mediated models are insensitive
to high scales in general and supergravity models are not.
The distinction here is that there  exists a nonvanishing $Str$ contribution
from a gauge-mediated model  as well in certain
cases when there is a   separation
of scales between fields which both contribute 
to the dynamical symmetry breaking sector, as occurs
for example in models with a dynamical messenger sector.
And there is always a nonvanishing $Str$ over the $T$ sector.

Having established that both the heavy and light
fields  in the DSB sector contribute to the squark scalar mass squared,
we see that the essential difference between
the first estimate for the squark mass squared (when $M_T \sim  M_{DSB}$)
and the case we are considering here is that the result
can be logarithmically enhanced, and furthermore that
we can establish the sign for the scalar mass squared if
it is indeed the logarithmically enhanced piece which dominates.

The dominant contribution in this case can be thought of
as arising in two steps. First, the scalar $T$ fields
get a  supersymmetry violating mass squared at two loops.
Notice that this mass is such that $Str(M_T)^2 \ne 0$.  The
$T$ scalar mass is then inserted into another two-loop diagram which generates
the mass squared for the squarks. The final contribution
can be obtained from the renormalization group equations
\beq
{d \delta M_{\tilde{T}}^2 \over d(log(\mu/M_{DSB})^2)}={ g_m^4 \over 256 \pi^4} (Str(M^2))_{DSB}C
\eeq
\beq  \label{above}
{d m_{\tilde{q}}^2 \over d(log(\mu/M_{DSB})^2)}= {g_s^4 \over 128 \pi^4} {4 \over 3} \delta M_{\tilde{T}}^2 (\mu)
\eeq
Here $C$ is a group theory factor. In Eq. \ref{above}, we
have taken the group theory factor for SU(3) explicitly, and have
used the fact that $Str(M^2)$ for one doublet flavor of the $T$'s is
$2M_1^2+2M_2^2-4M_T^2$, where $T_1$ and $T_2$ are the mass eigenstates. When integrating the above, we will be
interested in the result when $\mu=M_T$ for which the logarithm
is negative.
  The sign of $m_{\tilde{q}^2}$ depends on the
sign of $Str(M^2)$ in the DSB sector. If there is a dynamical
model with a separation of scales as the DSB sector and
the $Str$ is positive over the scales between
the two masses,   the mass squared for the squarks
and sleptons  will be positive. Furthermore the mass squared is  
  enhanced by the factor $Log (M_T/M_{DSB})^2$,
as can be seen by integrating the renormalization group equations.

It is an important requirement for  the mediator
model in the parameter regime  $M_m< M_T<M_{DSB}$  where  the 
logarithmically enhanced contributions to the squark
mass squared  dominate that the dynamical symmetry
breaking sector gives  positive $Str$. The
$T$ scalars  are then less massive than their  fermionic counterparts,
yielding {\it negative} $Str(M_T^2)$, which in turn
gives positive mass squared to the squarks.   If the only
contribution to the $T$ mass squared came from nonvanishing
$F$ terms, the $T$ scalars would be more massive than
the fermions and the squark mass squared would be negative.

It is essential that $Str M^2$ is not parametrically
larger    than $(F/S)$ from the light fields, since
the latter determines the gluino mass, whereas
the former determines the dominant contribution
to the squark mass. This turns out to be true of the models of Ref. \cite{pt},
and should be a generic feature of dynamical models
with a similar separation of scales. 

 We find that dynamical messenger
sectors work much more neatly in conjunction with a model
like this one, which gives the squark mass squared in two stages.
Were the messenger sector to give mass squared directly to the 
squarks,  it would generically be negative, though this
conclusion can perhaps be escaped in   specific models. 

It might also be considered an advantage of
this conjunction of the $T$ fields
with dynamical models that  the global
symmetry of the DSB sector does not need to be   sufficiently large to incorporate $SU(3)\times SU(2) \times U(1)$;
$G_m$ can be as small as SU(2). This  can sometimes permit a lower scale
of dynamical supersymmetry breaking, for which  higher
dimensional operators which yield   soft
scalar masses (which can be flavor changing)  are suppressed.


The remaining ingredient to consider  for this
class of models is  to establish the nonvanishing  messenger gaugino
mass.  Since the messenger gaugino mass
is getting a mass exactly as the gluino would in more direct
models of gauge-mediated DSB, but
without having to embed a group as large as the standard model,
 we know this is possible.
However, we wish to emphasize the special role of dynamical
models, and also to suggest that even models
in which the messenger gauge group is broken might work.
Discussions of two models can be found in an Appendix.

\section{ Intermediary Models}

Having considered a model in which it is gauge and not
superpotential interactions which communicate supersymmetry
breaking, we now consider a class of theories in which
 messenger gauge interactions are not necessary.   We
furthermore do  not require a complicated
mechanism to generate a nonvanishing singlet $F$-term.
The model does however incorporate singlets, or a dynamically
generated dimension four operator in the superpotential or the Kahler
potential.
  The    potential drawback to this model 
is the existence of additional local minima which do not
break supersymmetry or break standard model gauge groups.   Although
we  can argue (like in models \cite{dn,dns,dnns})
that these vacua are local minima, to make
the desired vacuum correspond to the deepest minimum
could require   additional structure \cite{we}.  
The advantage of the model is that the singlet field functions
very simply to produce an operator which permits direct
communication of supersymmetry breaking to messenger fields.

The field content of the model is as follows. First, we assume
the existence of a model which breaks supersymmetry dynamically.
In general, such models are chiral.  However,
we are interested in models which include at least one
vector representation, $V$, $\bar{V}$, of the gauge group whose dynamics
is responsible for breaking supersymmetry. The requirement
is that $V \bar{V}$ has  a nonvanishing $F$-term.
This can probably happen in many models. 

A specific example including a vector representation which
is worked out in the literature is in Ref. \cite{murayama}.
This model has the field content of the original ADS model
based on SO(10) with a singlet 16 \cite{ads5}, but has an additional H(10)
included. Murayama assumed this model can be analyzed perturbatively
(in reality there can be an unbroken strongly interacting gauge theory
so the analysis is not completely reliable \cite{pt2}) and
analyzed the vacuum. According to his result, the  invariant
$H^2$ has a nonvanishing  $F$-term for small mass.  The theory
could not be perturbatively analyzed with large mass.

Another example in which there are vector representations
is the model of Refs. \cite{iy,it}. Here the vacuum was not analyzed
for generic parameters, so it is not clear that the minimum has
nonvanishing $F$ term for one of the SU(2) mesons, though
it is possible.

In both of these examples, the vector field had nontrivial Yukawa
couplings in the superpotential. We suggest that the phenomenon
of nonvanishing $F$ components for a vector field might be much 
more generic. Suppose we take a model which breaks supersymmetry
through a combination of a superpotential generated by gaugino
condensation and perturbative terms which lift all flat directions.
Add to this model
a massive vector multiplet with the only tree-level
superpotential coupling being  $m_V V \bar{V}$,
which clearly lifts all new potential flat directions. This model
 clearly still breaks supersymmetry, since the old
equations of motion are still inconsistent.  This
can be seen most simply by first integrating out the massive
$V \bar{V}$ field. 

Now consider the full theory, including $V$ and $\bar{V}$.  We expect that
in general $(V \bar{V})_F$ has  a nonzero expectation value.
Notice that $V$ and $\bar{V}$ have nonvanishing expectation
values due to the dynamical term, and the mass term
removes any potential flat directions involving the $V$
fields.
If $V$ and $\bar{V}$ are heavy,   this $F$-term
  is mass suppressed. This can be seen by perturbing
the theory around the vacuum with vanishing $F$ term for the $V$ fields.
There is  a mass-suppressed tadpole for the $V$-field, which   induces a mass-suppressed $F$-term. The power dependence
of the mass-suppression is model-dependent.  If $V$ is light, the $V$ field
constitutes an approximate flat direction. We expect all vacuum
expectation values and $F$-terms to be governed by a combination
of its mass and the dynamical scale.   The
nonzero $F$ term is an assumption which can be verified in a detailed
analysis. A priori, there is no reason there should not be a comparable
$F$ term for $V$ as for other fields in the dynamical symmetry breaking
sector,  if $V$ is  light, and a mass-suppressed value if it is heavy. 
  A mass suppressed $F$-term should still be adequate, so long
as the mass is not too big and the $F$-term is nonzero.

It is not necessarily essential to have a mass term in models
with vector representations. The first two examples we
gave did not involve mass terms. What is important is
that all the flat directions are lifted. This is accomplished
through Yukawa interactions in the first two models mentioned.
In the absence of a specific model, the mass term is more generic,
but not necessarily essential.

 To complete the model, and transmit
supersymmetry breaking to the visible sector, we assume the existence of two singlet fields, $S$ and $\bar{S}$
and a vector-like messenger representation of the standard model or 
a GUT extension, 
$Q$ and $\bar{Q}$. The superpotential contains
$S V \bar{V}+M_S S \bar{S} + \bar{S} Q \bar{Q} + m_Q Q \bar{Q}$.
 Having two singlet field increases
the likelihood that  dynamical  interactions generate the necessary mass terms
and couplings and furthermore prevents potentially
dangerous interactions between the $V$ and $Q$ fields. 
Here we have not explained the absence of other terms
permitted by symmetries, but assume this will be explained
by a more fundamental theory. The fact that we want
all the couplings to be renormalizable (not suppressed by $M_{Pl}$)
is the reason we introduced  a vector-like representation into
the dynamical symmetry breaking sector. The large singlet mass
is necessary to separate the desired  minimum
in which supersymmetry is broken and field values
are independent of $M_S$ from the undesirable minimum
with large field values which grow with $M_S$. One
might hope that in a dynamical supersymmetry breaking model which has a singlet as an integral part of the model that no additional singlet
would be required. However, without an additional singlet, we
have found there is a new flat direction which destroys the
model as  a candidate for the supersymmetry breaking sector.
The mass for the $Q$ in the superpotential above  is required to keep the desired vacuum
stable.

The simplest way to analyze what happens is to first integrate
out the massive scalars. In this effective theory, there 
is a dimension four operator in the superpotential, $V \bar{V} Q \bar{Q}/M_S$.  If $V \bar{V}$ has a nonvanishing $F$ term,
and $Q$ has a mass, this model will work identically to the
usual gauge-mediated models.  No complicated couplings
are required in order to generate an $F$ term for the singlet.
At this point it should be clear that another possible model
just incorporates the dimension four operator directly. However,
if the scale is less than $M_{Pl}$, some explanation is warranted.
The operator could be generated by composite interactions,
but there would be common ``preons" to hidden and visible
sector fields, and one would also expect there to exist
fields which transform under the gauge groups of both sectors.
This can be dangerous, so we have instead considered the model
with singlets. An alternative possibility is that the necessary
operator arises from a Kahler potential coupling
involving the $Q$ and fields from a dynamical
sector, which is suppressed by a dynamical scale. Again, we have
yet to realize this possibility, and have therefore generated
the interaction via singlets.

It is readily seen in the effective theory that the old vacuum
is still a local mininum when $Q$ and $\bar{Q}$ vanish, since
all the old equations of motion for fields in the dynamical symmetry
breaking sector remain valid. However,  in the
absence of a mass term for $Q$, we see that the
mass term $Q \bar{Q}$ is suppressed by one power of $M_S$,
while the mass term for $Q Q^\dagger$ is suppressed by two powers,
so there would be an unstable field direction. This
is readily eliminated if the mass term for $Q$ is sufficiently large.
Our minimum is probably not the global minimum, which can
occur for field values of order $M_S$. Since the existing models
are based on local not global minima in any case we are no
worse off in this regard.

One can also analyze the vacuum in the theory with the $S$ and $\bar{S}$
fields still present. One finds similar conclusions, except that
 there is a nonvanishing vacuum expectation value of $S$ at order $1/M_S^2$.   This means that the vacuum we want is shifted slightly,
with the vacuum expectation value of fields shifting at order $1/M_S$.
 
 Here  we have found a simple generic mechanism to transmit
supersymmetry breaking to the visible sector.  We have
avoided the complications 
required to generate $F$-terms for the singlet.  It would
be nice to explicitly verify the nonvanishing $F$-terms
for the vector representation.   It
is not hard   to find a  theory with the necessary number
of vectors  which can be analyzed in a perturbative regime;
the analysis is however  numerically complicated.

\section{Mass Scales and Phenomenology}

The mediator models we have discussed  have a distinctive
mass spectrum which should be readily distinguishable
from that of other models of gauge-mediated supersymmetry breaking.
The intermediary models on the other hand give phenomenological
signatures very similar to models which have already been considered.
We briefly discuss the mass scales and consequences for our
models here and leave a more detailed analysis to further investigation.

We first discuss the  scale of supersymmetry breaking. This is
relevant because it will determine whether there will be
photonic decays inside the detector \cite{gam,dns,sl}.   This scale
also determines the importance of potentially flavor violating
scalar masses which can arise due to higher dimensional operators
in the Kahler potential, which can be significant when the dynamical
supersymmetry breaking scale is high.

In all cases, one anticipates a fairly high scale of supersymmetry
breaking in the mediator models.  This is because the
gaugino mass arises at three loops. If we constrain
the gluino mass to be about 100 GeV, this would imply
(with $\alpha_m\sim 1$) a supersymmetry breaking scale
at least  of order $10^6$ GeV, too high to be likely to
permit photon signatures. In a light
gluino scenario, this scale might be smaller by
an order of magnitude.  If the hidden sector
is provided by a dynamical supersymmetry breaking sector
with a hierarchy of mass scales, one expects an even
higher scale of supersymmetry breaking. For example,
in the model of Ref. \cite{pt} discussed in the Appendix,
if $N=5$,  the hierarchy of mass scales must be provided
by a small Yukawa coupling, $\tilde{\lambda}$. The scale of supersymmetry breaking is then higher by $1/\sqrt{\tilde{\lambda}}$.
If $N=7$, the scale of supersymmetry breaking is substantially
higher because the supersymmetry breaking communicated
to the messengers is suppressed by a ratio of mass scales.
It is readily seen that the scale of supersymmetry breaking
can be as high as $10^9-10^{10}$ GeV (where we have
 included the  logarithmic  enhancement of the gaugino mass).
This is quite high, but should be consistent in our models
in which the scalar masses are an order of magnitude higher
than gaugino masses (for like charges). The danger
is that there can be tree level flavor changing contributions
to the scalar masses; for the $N=7$ model they are suppressed
but interesting.
Clearly, the scale of supersymmetry breaking is model dependent.
 In all cases, it is likely to be large, but
the precise value depends strongly on the dynamical
symmetry breaking sector. The hierarchy
of mass scales present in the dynamical messenger
sector models (here the messengers are for the gauge group $G_m$)
is not an essential ingredient. The hierarchy played
two roles in Ref. \cite{pt}. First, it allowed for a calculable
model and second, it allowed for many states with standard
model gauge charge to be heavy. The calculability
of the model is not necessarily essential and potential
Landau poles are not a problem since it is $G_m$
and not $SU(3) \times SU(2) \times U(1)$ which is the gauged
global symmetry. However, a hierarchy of mass
scales between $M_{DSB}$ and $\sqrt{F_{DSB}}$ might
be a desirable feature in our models because it permits
a larger range for $M_T$. We therefore expect a high
scale for $F_{DSB}$, sufficiently high that photon signatures
will not occur, but sufficiently low that nonrenormalizable
terms are sufficiently small that FCNC effects are not too big.

The scale of supersymmetry breaking in the intermediary
models is determined by the unknown mass parameters.
So although supersymmetry breaking is transmitted  to the messengers at tree level,
the scale of supersymmetry breaking is likely to be high.
We conclude that    photon signatures are probably  not a  signature
for either type of model. 

The best tests of the models are the mass spectrum.
The mediator models imply that the gauginos which
are superpartners to the standard model gauge bosons
are the lightest of the superpartners (of like charge), with the scalars
of corresponding charges at least an order of magnitude
heavier. 

 A very light gluino scenario \cite{farrar} might still
be viable. However,  since $M_1$ and $M_2$ are also
very small, such a scenario is very constrained in light of LEP 1.5 (and LEP 2)
results, on top of any direct bounds
on the gluino itself.  In exploring the phenomenology of the light gluino
scenario of this model, it should be remembered that
in addition to the one-loop contributions to the gaugino masses
with intermediate standard model superpartners, there
is also the three-loop contribution we have already discussed.
If $\alpha_m$ is of order $1/(4\pi)$, these should be
of comparable importance. 
 
A less  restrictive parameter    range will occur if $\alpha_m$
is of order unity. In this case, one can have a phenomenologically
acceptable spectrum with the gaugino masses near their
current experimental limit.  
The upper bounds on masses arise from naturalness considerations
which would need to be redone for  the mediator models.
The lower limits are determined by recent LEP 1.5 results
and by the gluino mass limits. In models with a grand unified
mediator spectrum, the gaugino masses will
be related by the gauge couplings, and the
limit on any gaugino will imply limits for the other gauginos
as well.  The strongest bound on a gluino which is lighter than the squarks
comes from D0 \cite{d0} and is 144 GeV. This bound was
derived however for particular parameter choices.  The most
recent CDF \cite{cdf}
 publication gave results for various parameter choices; in some regions of parameters
there were no limits. UA2 gave a bound of 79 GeV \cite{ua2} which applied
for photino mass less than 20 GeV, which will be the case for
the simplest models.  It is conceivable there exists a gluino
window for heavier masses when one does not assume the
spectrum  of mediators is grand unified . Because this
can lead to other complications (a nonvanishing $D$-term for $U(1)_Y$
term for example), it is probably reasonable
to assume  the UA2  bound  applies. It would also
be worthwhile to extend the D0 bound to a broader
parameter range. The current gluino mass bound (assuming
the light gaugino scenario is not viable) is presumbably
on the order of 100 GeV. 

However, the stronger bound on parameters will come
from LEP 1.5  and LEP 2 results which will exceed the gluino mass
bound if GUT relations are assumed.  The most
recent published results \cite{lep2} indicate 
a bound on $M_2$ between about 20 and 30 GeV.
Because the sneutrino and slepton would be heavy
in this scenario, the bounds from chargino
and neutralino constraints can be readily interpreted
as a bound on the  gaugino mass matrix parameters.
The bound is strongly dependent on $\tan\beta$ and
favors a small value.  It also favors a negative and small value
for $\mu$. The naturalness of the model should be analyzed taking
into account
  the chargino and neutralino constraints which will
bound the parameters $\mu$ and $M_2$ and favor smaller
values of  $\tan\beta$  to maintain small $M_2$   in order to keep the
scalar spectrum not too heavy.

For a particular value of $\alpha_m$ and the gauge group $G_m$,
there will be a relation between the scalar masses and the
gaugino masses. This will however be model dependent as
there are in general contributions not only from the $F$ terms in the  dynamical
supersymmetry breaking sector, but from $Str(M^2)$ in the dynamical supersymmetry breaking
sector, whereas only the $F$
terms contribute to the gaugino mass. The qualitative
  prediction is that the matching conditions
for the scalar mass is on the order of ten times
bigger than in the standard phenomenological analysis
of gauge-mediated models. A more comprehensive study
of the viability and phenomenology of our parameter range
would be worthwhile.

The spectrum of the intermediary models is more similar
to that previously investigated, in Ref. \cite{sl} for example.
 The mass of the messenger quarks is however a free parameter,
unrelated to the dynamics of supersymmetry breaking
and its communication to the messenger squarks via the singlet.

\section{Generating \protect\boldmath$\mu$}

We have not yet addressed the other major stumbling block
to  gauge-mediated supersymmetry breaking, the generation of a $\mu$ term.
This is another place where models seem to be complicated.  The
solution to the $\mu$ problem can be the same as in previous
models of dynamical supersymmetry breaking. We present
another potential solution  which could also apply to other
models  and is really tangential to the rest of this letter.

A possible mechanism for generating a $\mu$ term is the following.
Introduce a singlet field $M$ and charged fields under an $SU(N_X)$
nonabelian gauge group $X(N_X)$ and $\bar{X}(\bar{N_X})$. Assume the
superpotential contains $M H_u H_d+ M X \bar{X}+M^3$ where $H_u$
and $H_d$ are the standard model Higgs superfields. Also assume
that $X$ and $\bar{X}$ have weak scale masses which were
generated in our model through $T_X$ fields, which
transform under the messenger gauge group and the group
under which $X$ transforms. (In  the
mediator models,  it would be best for the $T_X$ and $T$ masses
  arise from a common confinement scale so that the
mass of the $X$ fields is also weak scale. This can
happen  for example if $T$ and $T_X$ are composites of common preons
transforming under $SU(2)_m$ as well as preons
carrying either standard model of $SU(N_X)$ gauge charge).

We expect the $M$ scalar mass  to be zero at the matching scale (since it
is gauge-neutral) but  to
obtain a mass squared upon renormalization group running
(more quickly  for larger  $N_X$).   This
will scale the $M$ mass squared negative, and should lead to a VEV
for $M$ of order the weak scale, as in no-scale scenarios.

The $\mu B$ term is generated because the $M^3$ term implies
$\langle F_M \rangle \sim M^2$. If $\lambda \sim 1$, this is the correct
relation between $\mu$ and $\mu B$.
 This model is similar to that presented in Ref. \cite{dn}. The problem for
the simplest model there  was
that the superpotential, which only has dimension three operators,
preserves
  an $R$-symmetry (this was the U(1) identified in Ref. \cite{dn}) and
therefore contains an associated
 $R$-axion which couples to the $Z$. This problem is readily solved here
 because the $R$-symmetry is   anomalous with respect
to the gauge symmetry under which $X$ transforms, so the pseudoscalar
can be raised above the $Z$ mass through instanton-effects.   

\section{Conclusions}

The lack of elegance to visible sector models  is the chief reason
to view them with skepticism.
In this paper we have explored the question of whether
there can exist alternative mechanisms for communicating
supersymmetry breaking.   In the mediator models, 
   there is no need for singlets coupled directly to messenger quarks
or for a complicated superpotential, and furthermore,
there should be no new color or charge breaking minima. It should
be noted that we have assumed the $R$-axion intrinsic
to visible sector models is given a suitably high mass from supergravity
and cancellation of the cosmological constant \cite{bpr}.
No new superpotential couplings are required, aside
from that which gives the $T$ and $\bar{T}$ superfields a mass.
Furthermore, the global symmetry of the Lagrangian, which
is easy to obtain in many models of dynamical supersymmetry
breaking, is not necessarily   preserved by the dynamical supersymmetry
breaking vacuum.
Gauge interactions  and heavy fields
which transform under a messenger and standard model
gauge group  suffice to communicate supersymmetry
breaking.   However, for this class of models, there was
either a narrow range for $T$ mass or a dynamical
messenger gauge group was required. Furthermore,
the mass range which we predict will probably not
 produce as   natural a Higgs sector
as that  considered in  Ref. \cite{sl}, but should
be explored, and  is subject to experimental verification.

The intermediary models are much simpler as given.  Supersymmetry
breaking is directly communciated via a dimension four operator
in the superpotential. The massive singlet is necessary only
in order to generate this operator. There are several mass
parameters but they are fairly arbitrary, with the constraint
being only that the supersymmetry breaking scale should
not be so high that gravity-mediated supersymmetry
breaking is of comparable importance.  Nonetheless,
it would be of interest to produce these mass terms dynamically,
and work is underway here. It would furthermore be
of interest to verify the dynamical assumptions which were made;
   in particular,
it would be good to verify the nonvanishing $F$ terms for the
vectors.

Clearly, there is  much more
work to be done, in particular a better understanding
of the solution to the $\mu$ problem,   and a more
thorough investigation of the  phenomenology
of mediator models. Nonetheless, we  believe it
is quite interesting that there can be very different
scenarios for gauge-mediation of supersymmetry breaking. 
 
\section {Acknowledgements}
I thank Csaba Cs\'aki  and Witold Skiba
for conversations which   helped motivate
this work and Howard Georgi for pointing out a simplification and for
discussions,  and 
Bogdan Dobrescu, Ann Nelson, and Erich Poppitz for  very
valuable conversations.
I am also grateful 
  to Greg Anderson,   Michael Dine, 
  Asad Naqvi,    Michael Schmitt,   Sandip Trivedi,
and Darien Wood for  useful discussions. 

\section{Appendix: Generating \protect\boldmath$M_m$}

Generating $M_m$ is  not hard. However, analyzing
the vacuum in nonperturbative supersymmetry breaking models
is often stymied by the fact that the Kahler potential is not determined,
so the location of the vacuum is uncertain. For this
reason, it is often difficult to ascertain the global symmetry
group which remains at the minimum of the potential. We will
argue that it is not essential that the weakly gauged $G_m$
remains unbroken, although it is simplest to first consider
this possibility. The supersymmetry breaking mass for $G_m$ is   generated akin
to the mass of the true gauginos, through a diagram like
that of Figure~1.

We take the model analyzed in \cite{pt} as our first example.
This model might be 
 more complicated than necessary,
but since it has  already been completely analyzed,
it is a simple model for us to consider. 
Futhermore, it has the separation of scales
which could lead to the most
natural models of the type considered  in Section 2.
 The models considered in Ref.
\cite{pt}
 employed an $SU(N) \times SU(N-2)$ gauge group,
with   odd  $N$ and were shown to break supersymmetry. The couplings in the superpotential can
be chosen to preserve an  $SP(N-3)$. In order to incorporate
the standard model gauge group, it was necessary to take $N\ge 11$.
However, we only wish to incorporate the gauge group $G_m$ which
we take to be SU(2)
so we can take $N$
 as  small as 5,
although to make the theory perturbative without a small Yukawa
coupling prefers $N=7$.   

 This model establishes that
it is possible to generate supersymmetry breaking gaugino
masses for $G_m$ with a reasonable choice for the supersymmetry
breaking scale.   The above model is special in that one can determine the
low-energy vacuum and ascertain that a global symmetry
is preserved. The soft supersymmetry breaking mass
and $F$ type terms have been calculated, and it was established
that for the light fields that the singlet VEV scales as a parameter $v$
(related to more fundamental scales of the theory), the $F$ term
for the light fields scales as $v^2 (v/M)^{(N-5)}$ (or with a Yukawa
for $N=5$)), and that $Str(M^2)$ over the light fields scales
as $v^2(v/M)^{(N-5)}$. Here $M$ is likely to be $M_{Pl}$ but
could be some other scale.  Therefore the mass splittings and
$F$ type vevs contribute the same order of magnitude
to the squark and gaugino mass, aside from the logarithmic enhancements
discussed in the previous section. 
  
The next model we consider was given   by
   Affleck, Dine, and Seiberg \cite{ads52} 
and by   Dine, Nelson, and Shirman in Ref. \cite{dns}
and analyzed  by ter Veldhuis
in Ref. \cite{tatv}.    The model
has two 10's and two $\bar{5}$'s and a tree-level superpotential
\beq
W=10_1 \bar{5}_1 \bar{5}_2
\eeq
where this is the most general superpotential allowed by the symmetries.
 This superpotential preserves an SU(2) global symmetry (as
pointed out
in \cite{dns}) which  is spontaneously broken \cite{tatv}. In addition
to the supersymmetric mass for the SU(2) gauginos, there should also be   a supersymmetry breaking mass.
For example, both the $A$ and $F$ components of $10_1$ should
be nonzero at the minimum of the potential, so the diagram of Figure
1 should generate a supersymmetry breaking gaugino mass for the SU(2)
gauginos. Because of the supersymmetric contribution to their
mass, the   result for Figure ~3 will be different.
  However, no large ratio is expected for
values of the Yukawa coupling of order unity,
and furthermore, any ratio can be accommodated by adjusting the scale
of supersymmetry breaking.  

A model of this sort without a separation of scales
will require the $T$ mass to be of the same order of magnitude
as the mass of fields in the DSB sector. It is interesting
nonetheless 
  that a model with a broken global symmetry group can also
work.   The
  danger with such a model is that the  $D$-terms  can
be nonvanishing. This can potentially induce a VEV for
the $T$ or $\bar{T}$ field which would be a phenomenological disaster.
This will not occur   so long as $M_T^2>g^2 D_m$, where $D_m$
in this case is the $D$-term for SU(2) at the supersymmetry breaking minimum. The more serious danger for a nonvanishing $D_m$
is that the scalar mass will be generated  at tree level
or at one   loop,
making the gaugino to scalar mass ratio completely unacceptable.
Therefore we require vanishing (or loop suppressed) $D_m$.
This can follow because the gauge group is preserved,
or because there is a charge-conjugation or other symmetry which
protects the $D_m$ term. In the particular model just considered,
this requires that a custodial symmetry is preserved at some level,
which requires a small coupling ratio \cite{tatv}

      It  should be noted
that it is  not always true that there is a supersymmetry
breaking gaugino mass. For example,  in the $SU(6)\times U(1) \times U(1)$
model of Ref. \cite{dnns}, the messenger U(1) gaugino is massless.
However, in general, the U(1) gaugino should pick up a supersymmetry
breaking mass. A simple example of this is the $SU(3) \times SU(2) \times
U(1) $ model of Ref. \cite{ads1,dns}.  However in this case there 
is a  dangerous $D_m$ term.


\begin{thebibliography}{99}

\bibitem{bkn} T. Banks, D. B. Kaplan, A. E. Nelson, \pr{D49}{94}{779}.
 
 \bibitem{ads1} I. Affleck, M. Dine, N. Seiberg,  \np{B256}{85}{557}.
 


 \bibitem{gm}
M. Dine, W. Fischler, and M. Srednicki, \np{B189}{81}{575};
S. Dimopoulos and S. Raby, \np{B192}{81}{353}, \np{B219}{83}{479};
M. Dine and W. Fischler, \PL{B110}{82}{227}, \np{B204}{82}{346};
M. Dine and M. Srednicki, \np{B202}{82}{238};
L. Alvarez-Gaum\' e, M. Claudson, and M. Wise, \np{B207}{82}{96};
C. Nappi and B. Ovrut, \PL{B113}{82}{175}, H. P. Nilles, {\em Phys. Rep.}
{\bf 110} (1984) 1.

\bibitem{dn} M. Dine and A.E. Nelson, hep-ph/9303230, \pr{D48}{93}{1277}.

\bibitem{dns} M. Dine, A.E. Nelson, and Y. Shirman, hep-ph/9408384,
\pr{D51}{95}{1362}.

\bibitem{dnns}
 M. Dine, A.E. Nelson, Y. Nir, and Y. Shirman, hep-ph/9507378,
\pr{D53}{96}{2658};
M. Dine, Y. Nir, and Y. Shirman, hep-ph/9607397.


\bibitem{pt} E. Poppitz, S.P. Trivedi, hep-th/9609529


 \bibitem{we}I. Dasgupta, B.A. Dobrescu, and L. Randall, hep-ph/9507487,
N. Arkani-Hamed, C. D. Carone, L. J. Hall, H. Murayama,
\pr{D54}{96}{7032}.

\bibitem{martin} S. P. Martin, hep-ph/9608224.  

\bibitem{apt} G. Anderson, E. Poppitz, S. Trivedi, EFI-96-50, in progress.


\bibitem{mrm} N.
Arkani-Hamed,  J. March-Russell,
H. Murayama in progress.

\bibitem{yuk} E. Poppitz, Y. Shadmi and  S. Trivedi, \np{B480}{125}{1996},
hep-th/9605113; \pl{B388}{561}{1996}, hep-th/9606184;
C. Cs\'aki, L. Randall and W. Skiba, \np{B479}{65}{1996}, hep-th/9605108;
C. Cs\'aki, L. Randall, W. Skiba and R. Leigh, \pl{B387}{791}{1996},
hep-th/9607021.



 
 
\bibitem{bogdan} B. A. Dobrescu, hep-ph/9510424.

\bibitem{jmr} K. R. Dienes. C. Kolda, J. March-Russell, hep-ph/9601479.

\bibitem{murayama}  H. Murayama, \pl{B355}{95}{187}.

\bibitem{ads5} I. Affleck, M. Dine, N. Seiberg, \pl{B140}{84}{59}.


\bibitem{pt2} E. Poppitz, S. Trivedi, \pl{B365}{96}{125}.
 
\bibitem{iy} K.-I. Izawa, T. Yanagida, hep-ph/9602180,
{\it Prog. Theor.  Phys. 95}, 829, 1996.

 
\bibitem{it}K. Intriligator, S. Thomas, \np{B473}{96}{121}.
 
\bibitem{gam} P. Fayet, \pl{B84}{79}{421}.


\bibitem{sl} K. S. Babu, C. Kolda, and F. Wilczek, hep-ph/9605048,
S. Dimopoulos, S. Thomas, J. D. Wells,
hep-ph/9609434, J. Bagger, K. Matchev, D. Pierce, and R. Zhang, hep-ph/9609444.

\bibitem{farrar} G. Farrar, hep-ph/9612344 (and references therein).
 


\bibitem{d0} S. Abachi {\it et. al.}, \prll{75}{95}{618}.

\bibitem{cdf} F. Abe {\it et. al.} \prll{75}{95}{608}.

\bibitem{ua2} J. Alitti {\it et. al.}  \pl{B235}{90}{363}.

\bibitem{lep2} The ALEPH Collaboration,  \pl{B373}{96}{246}.   {\it.  Z Phys} {\bf C72} (1996) 549. 


\bibitem{bpr}J.A. Bagger, E. Poppitz, and L. Randall, hep-ph/9405345,
\np{B426}{94}{3}.


\bibitem{ads52}
  I. Affleck, M.  Dine, and  N. Seiberg, \prll{52}{84}{1677}.

\bibitem{tatv} T. A. ter Veldhuis, hep-th/9510121.




     




 \end{thebibliography}
\end{document}